\documentstyle[12pt]{article}
\input psfig.sty

\def\la{\mathrel{\mathpalette\fun <}}
\def\ga{\mathrel{\mathpalette\fun >}}
\def\fun#1#2{\lower3.6pt\vbox{\baselineskip0pt\lineskip.9pt
  \ialign{$\mathsurround=0pt#1\hfil##\hfil$\crcr#2\crcr\sim\crcr}}}

\begin{document}
\pagestyle{empty}

\begin{center}
\bigskip

\vspace{.2in}
{\Large \bf Dark Energy as a Modification\\
\medskip
of the Friedmann Equation}
\bigskip

\vspace{.2in}
Gia Dvali$^{1}$ and Michael S. Turner$^{2,3}$\\

\vspace{.2in}

{\it $^1$Center for Cosmology and Particle Physics\\
Department of Physics, New York University\\
New York, NY~~10003}
\vspace{0.1in}

{\it $^2$Departments of Astronomy \& Astrophysics and of Physics\\
Center for Cosmological Physics and Enrico Fermi Institute,\\
The University of Chicago, Chicago, IL~~60637-1433}\\

\vspace{0.1in}
{\it $^3$NASA/Fermilab Astrophysics Center\\
Fermi National Accelerator Laboratory, Batavia, IL~~60510-0500}\\

\end{center}

\vspace{.3in}
\centerline{\bf ABSTRACT}
\bigskip
Dark energy could actually be
the manifestation of a modification to the Friedmann equation
arising from new physics (e.g., extra dimensions).  Writing the
correction as $(1-\Omega_M)H^\alpha /H_0^{\alpha -2}$, we
explore the phenomenology and detectability of such.  We show
that:  (i) $\alpha$ must be $\la 1$; (ii) such a correction
behaves like dark energy with equation-of-state $w_{\rm eff}
= -1 + {\alpha \over 2}$ in the
recent past ($10^4> z\gg 1$) and $w=-1$ in the distant future and
can mimic $w<-1$ without violating the weak-energy condition;
(iii) $w_{\rm eff}$ changes, $dz/dw|_{z\sim 0.5} \sim {\cal O}(0.2)$,
which is likely detectable; and (iv)  a future supernova experiment like
SNAP that can determine $w$ with precision $\sigma_w$, could
determine $\alpha$ to precision $\sigma_\alpha \approx 2 \sigma_w$.

\newpage
\pagestyle{plain}
\setcounter{page}{1}
\newpage

\section{Introduction}

The discovery that the expansion of the Universe is speeding
up and not slowing down \cite{SCP,Hiz} has presented cosmologists and
particle physicists with a profound (and wonderful) puzzle.
In the context of general relativity this surprising result
can be accounted for by the existence of a smooth component
of energy with large negative pressure ($w\equiv p/\rho \la -1/2$),
dubbed dark energy, which accounts for about 2/3 of the critical
density \cite{dark_energy}.

A number of suggestions for the dark energy have been discussed including
quantum vacuum energy (cosmological constant), a very light and slowly evolving scalar
field, and a frustrated network of topological defects.
None is compelling and all have serious conceptual problems \cite{dark_energy}.

Another logical possibility is that the phenomenon of accelerated
expansion is actually a sign of a breakdown of the standard Friedmann equation
which governs the expansion rate, which
is the idea we explore here.

The high degree of isotropy and large-scale homogeneity observed in
the Universe implies that the metric of our 4-d spacetime can be
written in the Robertson -- Walker form with a single function --
the cosmic scale factor $R(t)$ -- describing the large-scale
dynamics of the Universe.  The issue then is the equation(s) that
govern the evolution of the cosmic scale factor.

It should be noted that the kinematics of the expansion -- acceleration or deceleration
-- can be discussed without regard to dynamics, and further,
that the current type Ia supernova (SNIa) data indicate a recent
period of acceleration ($\ddot R/HR^2 > 0$ for $z \la 0.5$) preceded
by an earlier period of deceleration ($\ddot R/H^2R <0$ for $z \ga
0.5$) \cite{turner_riess}.  Thus, simply allowing for modified
dynamics cannot eliminate the puzzling phenomenon of accelerated expansion.

In this paper, we investigate the addition to the Friedmann
equation of a term, $(1-\Omega_M)H^\alpha /H_0^{\alpha -2}$, which
can arise with theories with extra dimensions \cite{dgp}.  We know that at
early times the Friedmann equation is a good approximation
and this fact constrains $\alpha$ to be $\la 1$.  We further show that
such a modification has an equivalent description as dark
energy with time varying equation-of-state $w_{\rm eff}(z)$.  Finally,
we show that future supernova measurements envisioned with
SNAP \cite{SNAP} can constrain $\alpha$ to a precision of about half that
of $w$, i.e., $\sigma_\alpha \approx 2\sigma_w$, or
about $\sigma_\alpha \sim 0.1$.

In the next Section we discuss some theoretical motivations for
a modification to the Friedmann equation, and in the following Section
we discuss the cosmological phenomenology of ``$\alpha$ dark energy.''
We end with a brief summary.

\section{Motivations}

Both the hierarchy \cite{hierarchy1,hierarchy2}
and cosmological constant \cite{dgs} problems motivate
theories with large extra dimensions.
Extra dimensions that are either compact or have finite volume
manifest themselves exclusively at high
energies, above the compactification scale.
Such theories modify laws of gravity only at short distances,
below the size of the extra dimensions.
Consequently, all long-distance physics,
including the late-time cosmological evolution of the Universe,
is very close to the standard picture \cite{large_finite}.

In contrast, theories with {\it infinite volume}
extra dimensions \cite{dgp,dgs} modify the laws of gravity
in the far infrared, and at short distances (or early times)
the gravitational dynamics is very close to that of
the 4-dimensional Einstein gravity.
Consequently the  cosmological evolution is very close to
the standard FRW picture at early times, but gets modified at late times.
This modification can account for the observed accelerated expansion
of the Universe without dark energy; that is,
the  modified gravitational dynamics leads to
a ``self-accelerated'' Universe \cite{cd,ddg}.

For simplicity, we describe a model with a single extra dimension.
The effective, low-energy action takes the following  form
\cite {dgp}
\begin{equation}
{S}\,=\, {M_{\rm Pl}^2 \over r_c} \,\int\,d^4x\,dy\,\sqrt{g^{(5)}}\,{\cal R}\,+\,\int\, d^4x
\sqrt{g}\,\left ( M_{\rm Pl}^2\, R\,
+\,{\cal L}_{\rm SM}\right )\, .
\label{actD}
\end{equation}
where $M_{\rm Pl}^2 = 1/8\pi G$, $g^{(5)}_{AB}$ is
$5$-dimensional metric $(A,B=0,1,2,...,4)$, and $y$ is the extra
spatial coordinate.  The first term in Eq.~(\ref{actD}) is the
bulk $5$-dimensional Einstein action, and the second term is
the  $4$-dimensional Einstein localized on the brane (at $y=0$).
For simplicity we do not consider brane fluctuations.
The induced metric on the brane is given by
\begin{equation}
 g_{\mu\nu}(x)~\equiv~g_{\mu\nu}^{(5)}(x, y=0)\,.
\label{ind}
\end{equation}
The quantity $r_c$ is the crossover scale, the single new parameter.
It sets the scale beyond which the laws of 4-d gravity
breakdown and become 5-dimensional.

The existence of late-time, self-accelerated solutions can be seen
from Einstein's equations, obtained by the variation of
Eq.~(\ref{actD}) with an arbitrary 4-d matter source $T_{\mu\nu}$:
\begin{equation}
\label{Ee}
{1 \over r_c} {\cal G_{AB}} \, + \, \delta(y) \, \delta_A^{\mu}\delta_B^{\nu}\,
\left (\,  G_{\mu\nu} \, - \, 8\pi\, G \, T_{\mu\nu}\, \right )  \, = 0\, .
\end{equation}
Here ${\cal G_{AB}}$ and $G_{\mu\nu}$ are the 5-d and 4-d
Einstein tensors respectively.  Without the first term, Eq.~(\ref{Ee})
would be the standard Einstein equations for the induced 4-d metric $g_{\mu\nu}(x)$.
If we exclude the matter source ($T_{\mu\nu} \, =\, 0$),
then the only maximally-symmetric solutions to Eq.~(\ref{Ee})
are flat.  However, the presence of the first term
modifies this equation, though the modification becomes
significant only for very small values of the 4-d
curvature.  Consider the maximally symmetric FRW ansatz,
\begin{equation}
\label{s5}
ds_5^2 \, = \, f(y,H) ds_4^2 \, -\, dy^2,
\end{equation}
where $ds_4^2$ is the 4-dimensional maximally-symmetric metric,
and $H$ is the 4-d Hubble parameter, the second term
in Eq.~(\ref{Ee}) takes the standard form
$ G_{\mu\nu} \propto g_{\mu\nu} \, H^2$.

In the first term the $y$-derivatives of the warp-factor $f(y,H)$
must take care of the  $\delta(y)$-function, which leads to
an effective 4-d equation for $H$.  From simple  dimensional
arguments it follows that the first term must scale
as $\sqrt{H^2}$.  A calculation leads to Friedmann equation
for arbitrary 4-d brane-localized matter source $\rho_M(t)$ \cite{cd,ddg}:
\begin{equation}
\label{heq}
 H^2 \, \pm \, {H \over r_c}  \, = \, {8\pi G \rho_M \over 3}
\end{equation}
These general features persist for the arbitrary number of dimensions.

The basic point is that the higher-dimensional
action is suppressed relative to 4-dimensional one
by an inverse power of the crossover scale.
As a result, the scaling arguments suggest that
for the maximally symmetric ansatz the
higher-dimensional contribution should scale as
a lower power of $H^2$, and thus be important
at late times (small $H$).  This argument suggests
that infinite-volume extra-dimensional theories
have the potential to explain cosmic acceleration
with dark energy.

Before proceeding, let us make a few important points.
Naively, it appears that  as far as dark energy is
concerned the self-accelerated solutions of higher-dimensional
theory have the same number of new parameters
as the simplest model of dark energy, a cosmological constant. The
cosmological constant is replaced by the crossover scale $r_c$.
However, from the quantum field theory point of view there is a
crucial difference, the crossover scale $r_c$ is stable under
quantum corrections.

There is an interesting coincidence, which also motivates
the Hubble-scale value of $r_c$ and is unrelated to dark energy.
The scale $r_c$ also sets the distance at which
corrections (coming from higher-dimensional gravity) to
the usual metric for a gravitating source become important.
In particular, there are corrections to the
Schwarzschild metric \cite{ag}, which effect planetary motions.
The existing  phenomenological bounds on such deviations demand that the
crossover scale be large.  For instance, for Eq.~(\ref{actD})
the most stringent bound comes from lunar laser ranging
experiments that monitor the moon's perihelion precession with a great accuracy
and imply a lower bound for $r_c$
which is close to the  present cosmological horizon $H_0^{-1}$ \cite{dgz}.

 In the present paper we shall take a more radical and
generic attitude.   We shall use the notion of
infinite extra dimensions to motivate the
modification of Friedmann equation  at late times.
In that spirit, we shall assume that the physics that
modifies Friedmann equation satisfies the following
simple requirements:
\begin{itemize}

\item There is a single crossover scale $r_c$

\item To leading order, the corrections to Friedmann equation can
be parameterized a single term, $H^{\alpha}$

\end{itemize}

 These two assumptions fix the form of the modified Friedmann equation:
\begin{equation}
\label{hcor}
H^2 \, -  \, {H^{\alpha} \over r_c^{2 - \alpha}} \, = \, {8\pi \, G \, \rho_M \over 3}
\end{equation}
In order to eliminate the need for dark energy,
this term must specifically be:  $(1-\Omega_M)H^\alpha /H_0^{\alpha-2}$,
which implies that $r_c = (1-\Omega_M)^{1\over \alpha -2}H_0^{-1}$.

In the examples discussed above this stability of $r_c \sim H_0^{-1}$ is 
manifest (in contrast to the choice of a small vacuum energy).
It should remain true in our generalized analysis:
Since corrections to standard Einstein gravity are non-analytic
in $H^2$, they cannot be generated perturbatively in
the limit $1/r_c^{2 - \alpha} \rightarrow 0$.
Thus, the value of $1/r_c$ can only be
renormalized through {\it itself} and cannot be UV sensitive.
In other words, $1/r_c \, \sim \, H_0^{-1} $ plays the role of an
infrared cut-off for standard gravity, below which it is
replaced by a modified theory.

Finally, we would like to stress that we do not attempt to solve the
cosmological constant problem, i.e., the smallness of vacuum
energy. We simply postulate that it is zero due for whatever reason, and
study how the observed accelerated expansion could result from a
modification of the Friedmann equation in far infrared. In this respect
our approach is different (and complementary) to the idea of solving
cosmological constant problem by long-distance modification of
gravity \cite{dgs,nonlocal,tadpole}. In Refs.~\cite{nonlocal,tadpole} such
modification ensures that the vacuum energy (no matter how large)
gravitates extremely weakly, so that its over-all effect is reduced to the
one of a small cosmological constant {\it at all times}. Thus, the
resulting cosmology is indistinguishable from the standard FRW picture
with a small $\Lambda$-term, as opposed to the present case.

\section{Phenomenology of $\alpha$ Dark Energy}

Suppose that the effects of extra dimensions manifest themselves
as a modification to the Friedmann equation of the form
\begin{equation}
H^2 - (1-\Omega_M){ H^\alpha \over H_0^{\alpha -2}} = {8\pi G\rho\over 3} \,,
\label{1}
\end{equation}
where a flat universe has been assumed, $\rho$ is the total energy
density (today dominated by matter with $\Omega_M \approx 0.33$)
and $H_0$ is the present value
of the Hubble constant ($\approx 72\pm 7\,{\rm km\,sec^{-1}\,Mpc^{-1}}$).
The coefficient of the new term is fixed by requiring that it
eliminates the need for dark energy.

In the distant future, cosmic scale factor $R\gg 1$,
$$H \longrightarrow (1-\Omega_M)^{1\over 2-\alpha}H_0\,,$$
corresponding to asymptotic deSitter expansion.  For $\alpha = 0$
the new term always behaves just like a cosmological constant, while
for $\alpha =2$, the additional term corresponds to a ``renormalization''
of the Friedmann equation.

Equation (\ref{1}) can be recast in a more suggestive form
\begin{equation}
\left({H\over H_0}\right)^2 = 
(1-\Omega_M)\left({H\over H_0}\right)^\alpha + \Omega_M(1+z)^3 \,.
\label{2}
\end{equation}
Provided $\alpha <2$ (see below), the matter term will dominate
the r.h.s. for $10^4 \ga z \gg 1$, in which case $H \ \propto \ (1+z)^{3/2}$.
This implies that during the matter-dominated era
the $\alpha$ term varies as $(1+z)^{3\alpha /2}$,
which corresponds to an effective equation-of-state
$$w_{\rm eff} = -1+{\alpha \over 2}\qquad {\rm for\ } 10^4 \ga z \gg 1 \,.$$
[Recall, for constant equation-of-state $w\equiv p/\rho$, the energy density
varies as $(1+z)^{3(1+w)}$.]
During the earlier radiation-dominated epoch ($z\gg 10^4$),
where $\rho \propto \ (1+z)^4$ and $H\ \propto \ (1+z)^2$,
$\alpha$ dark energy has an effective equation of state
$w_{\rm eff} = -1 +{2\alpha \over 3}$ for $z\gg 10^4$.

To summarize, the effective equation-of-state of $\alpha$ dark energy varies
from $-1 +{2\alpha \over 3}$ during the radiation-dominated era
to $-1 + {\alpha \over 2}$ during the matter-dominated era to $-1$
in the distant future (see Fig.~1).  The change in the effective
equation-of-state is largest around $z\sim 0-1$, with $dw/dz \sim 0.2$.
This happens to be where SNe measurements of the expansion rate are
most sensitive to a variation in $w$ and a variation
of this size may well be detectable \cite{huterer_papers}.
It may be of some interest that for $\alpha < 0$, the effective equation-of-state
can be more negative than $w=-1$, without violating the weak-energy condition.

\begin{figure}
\centerline{\psfig{figure=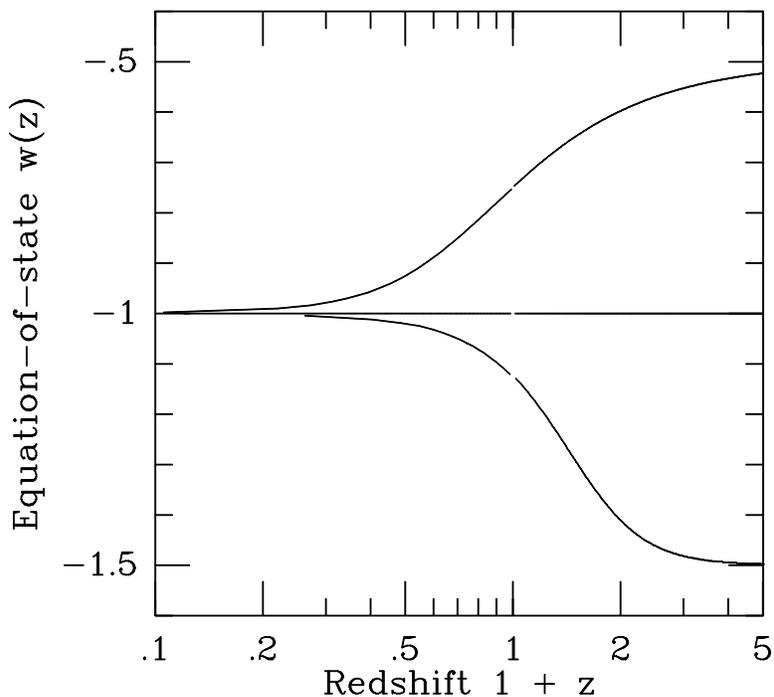,width=4in}}
\caption{Effective equation-of-state, $w_{\rm eff} \equiv -1 -{1\over 3}
{d\ln H^\alpha \over d\ln R}$, vs. $1+z$ for $\alpha
=-1,0,1$ (top to bottom).  Note, $1+z < 1$ corresponds
to the future, i.e., scale factor $R=1/(1+z) > 1$, where
$R=1$ today.
}
\end{figure}

To eliminate the need for dark energy, today the ``$\alpha$-term'' must be about
twice the matter term; however, to avoid interfering with the successful predictions of
the standard Friedmann equation, the $\alpha$ term must have been smaller
in the past.  In particular, the successful predictions of big-bang
nucleosynthesis set a limit to any new forms of energy density in the
Universe at $z\sim 10^{10}$ of less than a few percent of the standard
value (usually stated as limit to the number of neutrino species) \cite{bbn_limit}.
This in turn imposes an upper bound to $\alpha$:  

$$\alpha \la 1.95 \,.$$

A more stringent bound follows from requiring that the $\alpha$ term
not interfere with the formation of large-scale structure:  To achieve
the structure seen today from the primeval fluctuations whose imprint
was left on the cosmic microwave background requires a long matter-dominated
epoch \cite{TurnerWhite}.  In terms of $w$ this bound is:  $w\la -{1\over 2}$.
Requiring $w_{\rm eff} \la -{1\over 2}$ during the matter dominated
era leads to the bound:
$$\alpha \la 1 \,.$$

We remind the reader that these bounds apply only to modifications
of the Friedmann equation that aspire to explain dark energy,
i.e., where the corrections are significant today.  In finite-volume,
extra-dimensional models it is expected that $\alpha$ is larger
than 2 so that the corrections are most important at early
times.  For example, in many brane-world scenarios the modifications
effectively correspond to $\alpha = 4$ \cite{rhosquared}.

An expansion history (i.e., $H(z)$ vs. $z$) may be obtained
by picking a value for $H/H_0$ and solving for $z$.  From this,
the comoving distance as function of redshift $z$, $r(z) =
\int_0^z\, dz/H(z)$, can be computed.  Distance
vs. redshift diagrams [$r(z)$ vs. $z$] are shown in Fig.~2
(and in more detail in Fig.~3) for $\alpha = 1, 0, -1, -2, -3$.
Also shown are distance
vs. redshift diagrams for constant-$w$ dark-energy models.
From redshift $z=0$ to $z=2$ $\alpha$ dark energy can
be described approximately by a constant equation-of-state model with
$$w_{\rm eff} \approx -1 + 0.3 \alpha \,.$$

\begin{figure}
\centerline{\psfig{figure=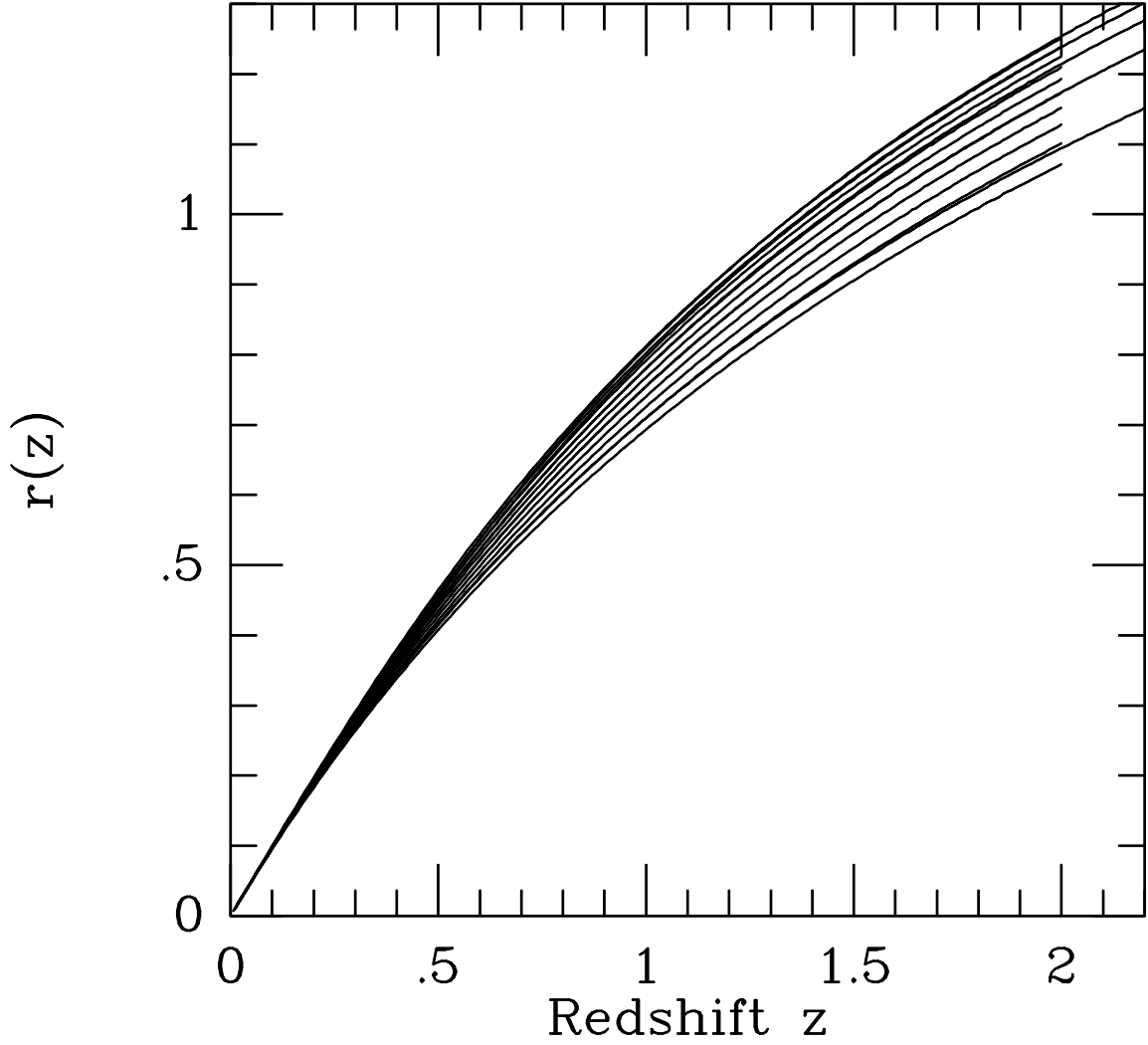,width=6in}}
\caption{Distance vs. redshift ($r(z)$ vs. $z$) for $\alpha = 1, 0, -1, -2, -3$
(full curves, from bottom to top) and for $w = -0.6, -0.7, -0.8, -0.9, -1.0,
-1.1, -1.2, -1.3, -1.4, -1.5$ (short curves, from bottom to top).
}
\end{figure}

\begin{figure}
\centerline{\psfig{figure=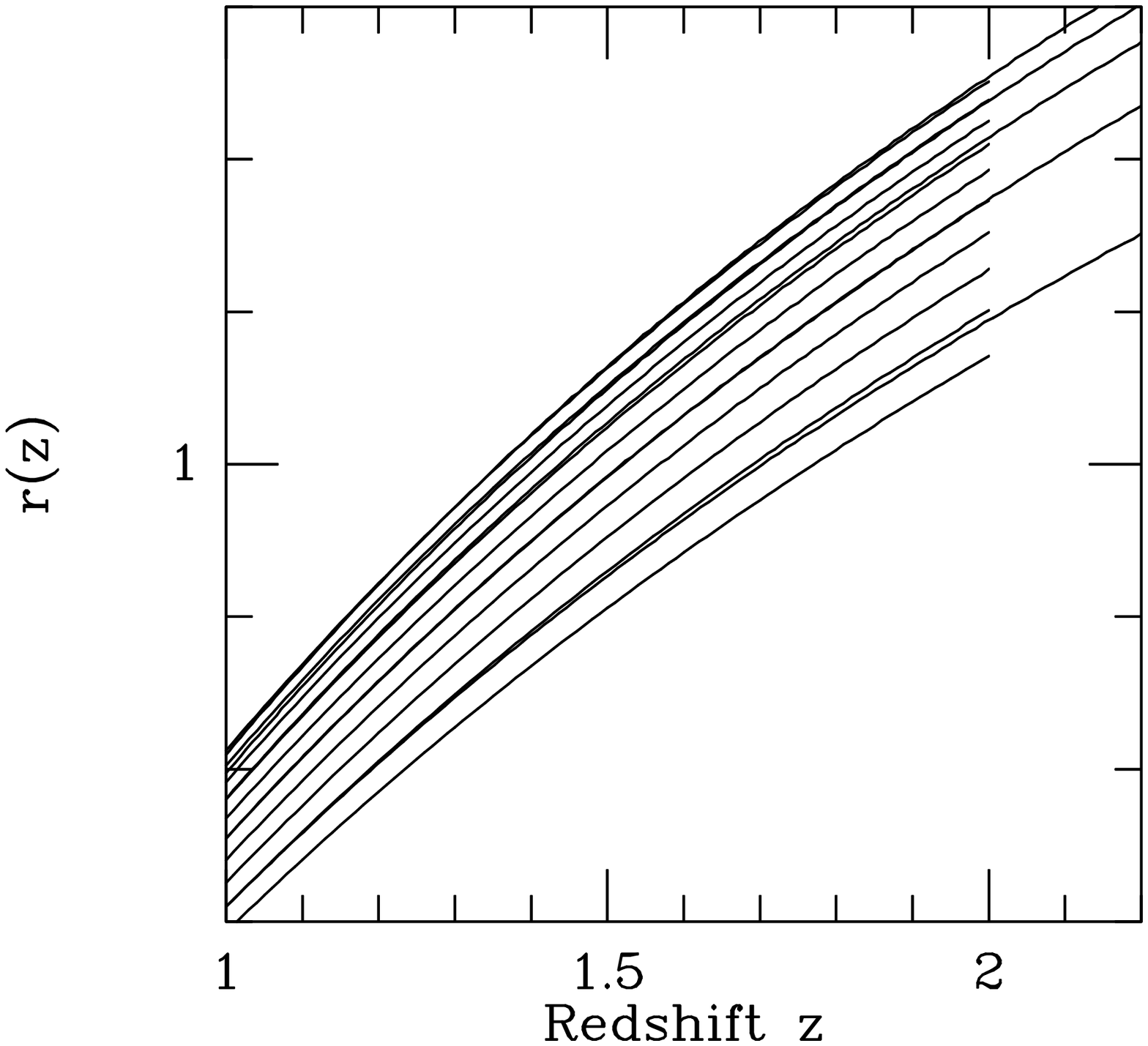,width=6in}}
\caption{Blow up of Fig.~2.
}
\end{figure}

An $\alpha$ dark energy model is completely described
by two parameters -- $\alpha$ and $\Omega_M$ -- and so
it is straightforward to address precisely the power of
supernovae observations to test it.  To this end, we have
constructed the Fisher matrix for a set of 2500
supernovae observations spanning $z = 0.2 - 1.7$
(like those that might be made by SNAP \cite{SNAP}):
\begin{eqnarray}
F_{ij} & = & \sum_k\,{1\over \sigma_k^2}\,w_i(z_k)w_j(z_k)\,,\nonumber \\
w_i (z) & \equiv & {\partial m(z) \over \partial p_i} \,,
\end{eqnarray}
where $m(z) \ \propto \ 5\log [r(z)]$ is the expected apparent magnitude
of a SNIa at redshift $z$, $\sigma_k$ is the expected
measurement accuracy (here assumed to be 0.15 mag),
and the parameters $p_i = \alpha$ and $\Omega_M$.
From the Fisher matrix we ``forecast'' the $1\sigma$ error ellipses
in the usual way.

Figure 4 compares the error ellipse in the $\alpha$--$\Omega_M$
plane with that in the $w$--$\Omega_M$ plane for constant-$w$
dark-energy models with $w=-1.3,-1,-0.7$.  What is most relevant
is the relative sizes of the error ellipses since the exact size
will depend upon details of the supernovae survey.  While the $\alpha = 1$
model ellipse aligns approximately with $w=-0.7$ as expected,
it is only about twice as large (not quite as large as expected
from the rough relationship between $w$ and $\alpha$).  Assuming
that SNAP can determine $w$ to around $\pm 0.05$, the same
observations could constrain $\alpha$ to around $\pm 0.1$,
sufficiently well to distinguish integer values of $\alpha$.

\begin{figure}
\centerline{\psfig{figure=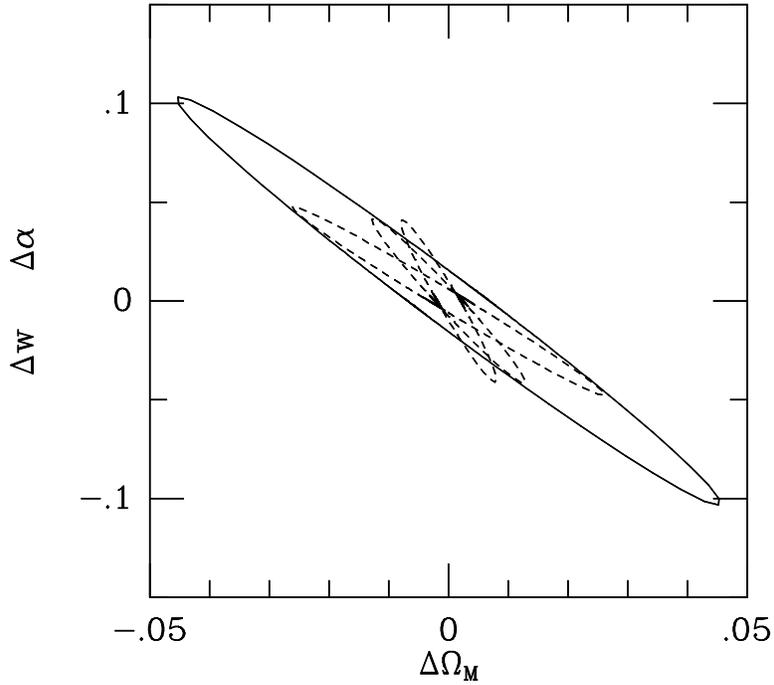,width=4in}}
\caption{Predicted error ellipses in the $\alpha$ -- $\Omega_M$
and $w$ -- $\Omega_M$ planes for $w = -1.3, -1.0, -0.7$ (moving
counterclockwise) and for $\alpha = 1$ for a SNAP-like supernova
experiment, assuming $\Omega_M=0.33$.
}
\end{figure}

\section{Concluding Remarks}

We have argued that dark energy may actually be a modification to
the standard Friedmann equation arising from new physics (e.g., infinite-volume
extra dimensions).  Writing the modification as
$(1-\Omega_M)H^\alpha/H_0^{\alpha -2}$, the
cosmological phenomenology can be summarized by:

\begin{itemize}

\item{} The effective equation-of-state of $\alpha$ dark energy evolves
from $-1+{2\alpha \over 3}$ ($z\gg 10^4$) to
$-1+{\alpha\over 2}$ $(10^4 \ga z\gg 1)$ and asymptotically
to $-1$ in the distant future ($R\rightarrow \infty$ or $z\rightarrow
-1$).

\item{} The maximal change in $w_{\rm eff}$ occurs recently,
$z\sim 0 -1$, with $(dw/dz)_{z\sim 0.5} \approx 0.2$.
This is where measurements by a future experiment like SNAP are most
sensitive and could probably detect a variation of this magnitude.

\item{}  For $\alpha < 0$, the effective equation-of-state $w_{\rm eff}$ 
can be less than $-1$ without violating the weak-energy condition.

\item{}  Around $z\sim 0-2$, an $\alpha$ dark-energy model can
be described approximately by constant $w$ with
$w\approx -1 + 0.3\alpha$.

\item{}  If SNAP can measure $w$ to a precision $\sigma_w \approx 0.05$,
it could determine $\alpha$ to about twice that, $\sigma_\alpha \approx
2\sigma_w \approx 0.1$.

\end{itemize}

If the phenomenon of dark energy is associated with corrections
to the Friedmann equations from new physics and
involves corrections like $H^\alpha$, it seems likely that future
supernova measurements could test this hypothesis and determine
$\alpha$ to a precision of order 10\%.  Thus, $\alpha$ dark
energy is not only intriguing, but it is also testable.

\paragraph{Acknowledgments.}
This work was supported by
the DoE (at Chicago and at Fermilab), by the NASA (at Fermilab
by grant NAG 5-7092), by GD's David and Lucille
Packard Foundation Fellowship for  Science and Engineering
Alfred P. Sloan Foundation Research Fellowship, and by
the NSF (grant PHY-0070787 at NYU).

\end{document}